# Frequency Adapted Phase Transition of Interacting Nano-Magnetic Ensemble

*Nilangshu K. Das[*] and P. Barat*

Variable Energy Cyclotron Center, 1/AF, Bidhan Nagar, Kolkata-700064, India

*e-mail: niludas@vecc.gov.in

**Each single domain nano-magnet acts as a magnetic dipole in addition it oscillates its magnetization about the easy axis and rotates coherently depending upon temperature and anisotropy. In an ensemble of nano-magnets, the relaxation time of a nano-magnet tunes with the long range dipolar interaction which in turn is determined by the particle size, density and the number of nano-magnets present in the ensemble. Hence, the aggregation of interacting nano-magnetic dipoles, demonstrates both experimentally and theoretically as a model system to detect intriguing co-operative physical phenomena. Here we show a new variant of phase transition from paramagnetic to diamagnetic phase by changing the frequency of the applied sinusoidal magnetic field for a nano-magnetic ensemble. This phenomenon unravels a new insight of physics and it may be significant on the design and development of magnetic devices. The simulation of the system is well in agreement with the experimental results.**

The dipolar nano-magnetic ensemble appears in many contexts and several theories have been suggested to understand their behavior due to applied magnetic field[1-5]. The deviation of the behavioral patterns of interacting nano-magnetic particles from the Stoner-Wohlfarth[6] and Neel-Brown[7] models are accredited to dipolar interaction and applied magnetic field. In the nano-magnetic particle system dipolar interaction is ubiquitous due to the long range interactions and manifests many exotic physical phenomena[8]. The dynamic magnetization of a dipolar nano-magnetic ensemble due to the applied magnetic field is dictated by anisotropic energy, dipolar interaction energy and Zeeman energy. In most of the references the dynamics of the nano-magnets are studied while the Zeeman energy supersedes the other two forms of energies[9,10] but to appreciate the synergetic effect of these three energies their magnitudes should be comparable. The relaxation time of non-interacting nano-magnetic particles depends on the anisotropic energy barrier only, however the presence of dipolar interaction has a predominant influence in determining the relaxation time and has been addressed extensively at several theoretical[11] and experimental studies[12]. These theoretical approaches are primarily based on the mean field theory, where an effective interaction field appears to be responsible for the observed effects[13] and admits analytical solutions. Traditional mean field theories overlook local fluctuations due to dipolar interaction at the time of delineating bulk-continuum. Therefore the mean field approach has its short comings in its effort to explain the exact behavioral pattern of magnetization for an applied magnetic field on a finite assembly of nano-magnetic particles. In the interacting nano-magnetic particle system local interaction field brings the system into a stable minimum energy state[14-16]. External magnetic field acts as a perturbation to this interacting particle system and disturbs the magnetic ordering and the stability in turn. The system will try intrinsically to oppose the applied field, the cause of its instability. This originates a new concept of diamagnetism in the interacting nano-magnetic ensemble.

In this letter, we report an exciting dynamic feature of nano-magnetic ensemble under the influence of spatial variation of interaction energy along with a temporal variation of Zeeman energy. We focus on the dynamics of interacting nano-magnets due to the influence of sinusoidal magnetic field. The switching probability of each nano-magnet at different spatial locations in the ensemble is different and hence the overall magnetization is time reliant. So the magnetization of the ensemble must have a phase lag with respect to the applied magnetic field[5]. On this account we have taken 15x15x15 Ising-like Cobalt nano-particles of dimension 8.5 nm arranged in a cubic array of lattice constant 11 nm. For simplicity of computation the easy axes of the nano-magnets are chosen to be parallel to the z-axis. The computation is carried out at 300K, well below the magnetic ordering temperature[14] and hence the initial ordered state is taken to be antiferromagnetic[15]. In this computation the time of switching is assumed to be instantaneous and the state of the magnetic moment of the nano-magnets are taken to be either parallel or anti-parallel to the direction of the applied oscillatory magnetic field which is along the z-axis. The magnetic field in the z-direction at any arbitrary j[th] lattice location is given by equation (1) where $s^i = \pm 1$ is the state of the i[th] nano-magnet and the other variables have their usual meaning.

$$\boldsymbol{H_T}(z^j, f, t) = H_a \sin 2\pi f t \, \hat{\boldsymbol{z}} + M_s V \sum_{i \neq j}^{N} s^i \frac{1}{r_{ij}^3} \left( \frac{3(z^i - z^j)^2}{r_{ij}^2} - 1 \right) \hat{\boldsymbol{z}} \quad (1)$$

and the corresponding relaxation time is

$$\tau(z^j, f, t) = \tau_o \exp\left\{ \frac{KV}{k_B T} \left( 1 + S^j \frac{H_T(z^j, f, t)}{H_K} \right)^2 \right\} \text{ where } H_K = \frac{2K}{M_S} \quad (2)$$

In the simulation, the volume of the nano-magnetic particles, lattice constant and applied magnetic field are so chosen that the three forms of energies are of the order of $10^{-20}$ Joules. The dipolar field at each lattice site in the ensemble depends upon the overall orientation of the particles and it changes whenever a single nano-magnetic particle switches from one state to the other. Therefore, in the process of simulation the dipolar field is computed for each nano-magnetic particle whenever switching occurs at any lattice site. The whole process is stochastic and the particle switches according to the probability derived from its relaxation time. The simulation is carried out at every nanosecond interval and the magnetization was calculated from the average values of the calculated magnetization over ten periods of the applied magnetic field. The frequency is varied from 50 kHz to 10 MHz and the susceptibility of the system is plotted with respect to the frequency as shown in Figure 1. The simulation result shows that the susceptibility goes to a diamagnetic phase for certain frequency.

This study inspired us to verify the findings experimentally. Nano particles of Co-Silica and Ni-Silica are prepared by conventional *sol-gel* technique in a controlled way to obtain the particle size in the range of 5 to 10 nm with narrow distribution. It was characterized by Scanning Electron Microscopy (SEM) and Transmission Electron Microscopy (TEM). SEM micrograph reveals the average size of the silicon substrate is around 10 μm. TEM micrograph reveals the existence of the crystallinity of the nano participles by the selected area diffraction. From TEM images of Cobalt and Nickel nano particles we concluded the average distance among the nano particles embedded in the silicon matrix is more than 10 nm. The sample susceptibility has been measured by a self-inductance type absolute coil susceptometer[18] at 300K. Typical results are shown in Figure 2. A transition from paramagnetic to diamagnetic phase with frequency is clearly evident from the figures.

Observation of phase transition from paramagnetism to ferromagnetism is quite common in literature however the phase transition from paramagnetism to diamagnetism is rare[19]. We have observed a paramagnetic to diamagnetic phase transition in an ensemble of coupled bi-stable nano-magnetic dipole oscillators subjected to a periodic external magnetic field. Classically the ensemble can be looked as a coupled dynamical system with varying internal delays. These internal delays are *incognito* the relaxation time of the nano-magnets. Nano-magnetic ensemble tries to orient themselves parallel to the applied magnetic field. This alignment process of the dipoles in the ensemble is of dual steps, an instigation of switching at the surface by the applied field and secondly, a cooperative switching. The switching of the nano-magnets begins at the surface because in finite ensemble the dipolar interaction field falls significantly at the surface[20] with respect to the core. This switching of nano-magnets at the surface sets off a series of switching in the ensemble due to the sudden change in interaction field caused by the neighbor reversal[21]. Gradually this alignment process penetrates into the sub-surface *i.e.,* the dipolar orientation in the outer shell is aligned with respect to the applied field whereas the inner core retains the state of opposition. Thus in addition to the external field the core also experiences an opposing field generated collectively by the dipoles of the shell. The thickness of the shell increases in due course of time and the opposing field by the shell becomes sufficiently high to trigger a cooperative switching. The temporal location of this cooperative switching with respect to the applied field determines the degree of paramagnetism or diamagnetism of the ensemble as shown in Figure 3.

A critical field is required to overcome the anisotropic barrier and dipolar potential of the nano-magnet at the surface to initiate switching. This can happen when $H_T = H_K$. The time to initiate this reversal of magnetization from the surface is given by the equation (3) where $H_{dd-surface}$ is the dipolar field experienced by the nano-magnets at the surface.

$$t_0 = \frac{T}{2\pi}\sin^{-1}\frac{H_K - H_{dd-surface}}{H_a} \quad (where\ T = \frac{1}{f}) \qquad (3)$$

After $t_0$ this reversal of magnetization creeps into the sub-surface of the ensemble until a critical shell thickness develops to initiate the cooperative switching in the core. The system inherently takes time, $t_d$ to develop the critical shell thickness. Consequently the cooperative switching is further delayed by $t_d$ over $t_0$, where $t_d$ depends on relaxation time of the nano-magnets. The transition time of total reversal of population of nano-magnets, $t_r = t_0 + t_d$. This $t_r$ is the governing factor of the state of the overall magnetization. The maximum value of $t_0$ is $T/4$ when $H_a = (H_K - H_{dd-surface})$. In that occation $t_r$ is always greater than $T/4$ and the ensemble should show diamagnetism for all frequencies. If $H_a > (H_K - H_{dd-surface})$, $t_0$ is less than $T/4$, and the contribution of $t_d$ becomes important to make $t_r > T/4$. This can be achieved by increasing the frequency or by increasing relaxation time of the nano-magnets or the combination of both.

This novel and original observation has immense possibilities to be explored in the field of nano-magnetism in particular and in physics at large.